\documentclass[letterpaper,11pt]{article}

\usepackage[top=1in, bottom=1in, left=1in, right=1in]{geometry}  

\usepackage{amsmath, amsthm}  
\usepackage{mathtools}   
\usepackage{bm}
\usepackage{amsfonts}

\usepackage{graphicx} 
\usepackage{enumerate}
\usepackage[authoryear]{natbib}
\usepackage{url} 

\usepackage{enumitem}   
\usepackage[table,xcdraw,dvipsnames]{xcolor}   
\usepackage{pdflscape}
\usepackage{afterpage}
\usepackage{capt-of}
\usepackage{adjustbox}

\usepackage[many]{tcolorbox}
\usepackage{multirow}   
\usepackage{tabularx} 

\usepackage{float}   
\usepackage{nameref}   

\usepackage[font=normalsize]{caption}  
\usepackage{subcaption}  
\usepackage[titletoc,title]{appendix}  
\usepackage{longtable} 

\usepackage[final]{hyperref} 
\hypersetup{
	colorlinks=true,       
	linkcolor=blue,        
	citecolor=blue,        
	filecolor=magenta,     
	urlcolor=blue         
}

\newcommand{\utwi}[1]{\mbox{\boldmath $ #1$}}

\newcommand{\bA}{{\utwi{A}}}

\newcommand{\bE}{{\utwi{E}}}
\newcommand{\bF}{{\utwi{F}}}

\newcommand{\bH}{{\utwi{H}}}
\newcommand{\bI}{{\utwi{I}}}

\newcommand{\bM}{{\utwi{M}}}

\newcommand{\bQ}{{\utwi{Q}}}

\newcommand{\bW}{{\utwi{W}}}
\newcommand{\bX}{{\utwi{X}}}

\newcommand{\bZ}{{\utwi{Z}}}

\newcommand{\bPhi}{{\utwi{\mathnormal\Phi}}}

\newcommand{\bepsilon}{{\utwi{\epsilon}}}
\newcommand{\bGamma}{{\utwi{\Gamma}}}

\newcommand{\bOmega}{{\utwi{\Omega}}}
\renewcommand{\bPhi}{{\utwi{\Phi}}}

\newcommand{\calM}{{\mathcal M}}
\newcommand{\calN}{{\mathcal N}}

\def\spacingset#1{\renewcommand{\baselinestretch}%
{#1}\small\normalsize} \spacingset{1}

\newtheorem{remarkx}{Remark}

\newtheorem{condx}{Condition}

\DeclarePairedDelimiterX{\norm}[1]{\lVert}{\rVert}{#1}   

\newcounter{question}

\renewcommand{\hat}{\widehat}

\newcommand{\mybib}{C:/Users/chloe/Dropbox/papers/dynamic_networks}
\renewcommand{\baselinestretch}{1.2} 
\graphicspath{ {images/} }

\title{Factor Model for High-Dimensional Dynamic Networks: with Application to International Trade Flow Time Series $1981$--$2015$}
\author{
	Yi Chen and Rong Chen
}
\date{\today}

\begin{document}
\maketitle

\begin{abstract}

Dynamic network analysis has found an increasing interest in the literature because of the importance of different kinds of dynamic social networks, biological networks and economic networks. Most available probability and statistical models for dynamic network data are deduced from random graph theory where the networks are characterized on the node and edge level. They are often very restrictive for applications and unscalable to high-dimensional dynamic network data which is very common nowadays. In this paper, we take a different perspective: The evolving sequence of networks are treated as a time series of network matrices. We adopt a matrix factor model where the observed surface dynamic network is assume to be driven by a latent dynamic network with lower dimensions. The linear relationship between the surface network and the latent network is characterized by unknown but deterministic loading matrices. The latent network and the corresponding loadings are estimated via an eigenanalysis of a positive definite matrix constructed from the auto-cross-covariances of the network times series, thus capturing the dynamics presenting in the network. The proposed method is able to unveil the latent dynamic structure and achieve the objective of dimension reduction. Different from other dynamic network analytical methods that build on latent variables, our approach impose neither any distributional assumptions on the underlying network nor any parametric forms on its covariance function. The latent network is learned directly from the data with little subjective input. We applied the proposed method to the monthly international trade flow data from 1981 to 2015. The results unveil an interesting evolution of the latent trading network and the relations between the latent entities and the countries. 

\vspace{2em}

\end{abstract}

\section{Introduction}
Nowadays in a variety of fields, such as economics and social studies, researchers observe high-dimensional matrix-variate time series where observation at each time point is in a matrix structure. One special kind of such data is a time series of square matrices that describe pairwise relationships among a set of entities. For example, international trade commodity flow data between $n$ countries over a period of time can be represented as matrix time series $\{ \bX_t \}_{t=1:T}$, where $\bX_t$ is a $n \times n$ matrix, and each element $x_{ij,t}$ is the directed level of trade from nation $i$ to nation $j$ at time $t$. The $i$-th row represents data for which nation $i$ is the exporter and the column $j$ represents data for which $j$ is the importer. Since the data are based on pairs of nations, the diagonal representing the relationships of nations with themselves is generally absent. 

Such \emph{network/relational data} that consist of measurements made on pairs of entities have been researched from various aspects over the last decades. Developed statistical models for \emph{static network analysis} include the class of exponential random graph models (ERGMs) that are analogous to standard regression models (\cite{wasserman1996logit, robins2007introduction, lusher2012exponential}), the class of stochastic block models that are, in their most basic form, essentially a mixture of classical random graph models (\cite{holland1983stochastic, nowicki2001estimation, daudin2008mixture, airoldi2008mixed, karrer2011stochastic}) and the class of latent network models that use both observed and unobserved variables in modeling the presence or absence of network edges (\cite{hoff2002latent,hoff2008modeling,hoff2005bilinear,hoff2009multiplicative, hoff2015multilinear, hoff2015dyadic, cranmer2016navigating}). All models mentioned thus far only consider a `snapshot' $\bX_t$ of a dynamic process in a given `slice' of time $t$, and thus not able to discover the dynamic pattern of the network nor to answer scientific questions concerned with the evolution of networks over time.

Statistical research on \emph{dynamic network analysis} is less developed compared to the existing literature on modeling of static network graphs. While there has been a substantial amount of work done in the past decades on the mathematical and probabilistic modeling of dynamic processes on network graphs (see \cite{barrat2008dynamical}), there has been comparatively much less work on the statistical side. Snijders and colleagues (\cite{snijders2001statistical, huisman2003statistical,snijders2005models,snijders2006statistical,snijders2007modeling, snijders2010introduction, snijders2010maximum}) developed an actor-based model for network evolution that incorporated individual level attributes. The approach is based on an economic model of rational choice, whereby actors make decisions to maximize individual utility functions. \cite{hanneke2010discrete} and \cite{krivitsky2014separable} introduced a class of temporal exponential random graph models for logitudinal network data (i.e. the networks are observed in panels). They model the formation and dissolution of edges in a separable fashion, assuming an exponential family model for the transition probability from a network at time $t$ to a network at time $t+1$. \cite{westveld2011mixed} represent the network and temporal dependencies with a random effects model, resulting in a stochastic process defined by a set of stationary covariance matrices.  \cite{xing2010state} extends an ealier work on a mixed membership stochastic block model for static network (\cite{airoldi2008mixed}) to the dynamic senario by using a state-space model where the mixed memembership is characterized through the observation function and the dynamics of the latent `tomographic' states are defined by the state function. Estimation is based on the maximum likelihood principal using a variational EM algorithm. They deduced from random graph theory \cite{crane2016dynamic, krivitsky2014separable}. These methods are deduced from random graph theory and model the relational data at relation (edge) or entity (node) level, and thus often confronted with computational challenges, overparametrization, and overfitting issues when dealing with high-dimensional matrix time series, which are very common in economics and social networks nowadays.   

In contrast to the pre-existing research in dynamic network analysis, the approach we propose is more time series oriented, in that a dynamic network is treated as a time series of matrix observations -- the relational matrices-- instead of the traditional nodes and edges characterization. We adopt a matrix factor model where the observed surface dynamic network is assume to be driven by a latent dynamic network with lower dimensions. The linear relationship between the surface network and the latent network is characterized by unknown but deterministic loading matrices. The latent network and the corresponding loadings are estimated via an eigenanalysis of a positive definite matrix constructed from the auto-cross-covariances of the network times series, thus capturing the dynamics presenting in the network. Since the dimension of the latent network is typically small or at least much smaller than the surface network, the proposed model often result in a concise description of the whole network series, achieving the objective of dimension reduction. The resulting latent network of much smaller dimensions can also be used for downstream microscope analysis of the dynamic network. 

Different from \cite{xing2010state} that summarize the relational data by the relationships between a small number of groups, we impose neither any distributional assumptions on the underlying network nor any parametric forms on its covariance function. The latent network is learned directly from the data with little subjective input. The meaning of the nodes of the latent network in our model is automatically learned from the data and is not confined to the `groups' to which the actors belong, which provide a more flexible interpretation of the data. Additionally, our modeling framework is very flexible and extendable: Using a matrix factor model framework, it can accommodate continuous and ordinal relational data. It can be extended to incorporate prior information on the network structure or include exogenous and endogenous covariate as explanatory variables of the relatioships. 


The remaining part of the paper is organized as follows. In Section 2, we introduce two factor models for network time series data and discuss their interpretations. In Section 3, we present an estimation procedure and the theoretical properties on the estimators. In Section 4, we study and compare the finite sample properties of the two proposed models on synthetic datasets. In Section 5, we apply the proposed factor models to the international trade flow time series from 1981 to 2015.  

\section{Factor Model for Dynamic Networks}

Let $\bX_t$ represent the $n$ by $n$ matrix of observed pairwise asymmetrical relationships at time $t$, $t = 1, \ldots, T$. A general entry of $\bX_t$, denoted as $x_{ij,t}$, represents the directed relationship of object $i$ to object $j$. For example, in international trade context $x_{ijt}$ represents the import/export from country $i$ to country $j$ at time $t$. And in the transportation context $x_{ijt}$ represents the a trip from location $i$ to location $j$ starting at time $t$. 

Our model for asymmetric directional matrix time series can be written as:
\begin{equation} \label{eqn:fac_AA}
	\bX_t = \bA \bF_t \bA' + \bE_t, 
\end{equation}  
where $\bA$ is an $n \times r$ (vertical) matrix of "loadings" of the $n$ objects on a relatively few $(r < n)$ aspects or types of objects. $\bF_t$ is a small, usually asymmetric, $r$ by $r$ matrix giving the directional relationships among the basic $r$ types, and $\bE$ is simply a matrix of error terms. $\bA$ relates the observed entities to the latent entities and $\bF_t$ describes the interrelations among the latent entities. 

An interesting feature of the above model is that while $\bF_t$ is allowed to be asymmetric, the left and right loading matrices $\bA$ are still required to be identical. This provides a description of data in terms of asymmetric relations among a \textit{single} set of entities, rather than envisioning a different set of aspects for the countries in their "importing" role than they have in their "exporting" role. A second possible apprach, where the left $\bA$ may be different from the right $\bA$, can be written as:
\begin{equation} \label{eqn:fac_A1A2}
\bX_t = \bA_1 \bF_t \bA_2' + \bE_t, 
\end{equation}  
where $\bA_1$ and $\bA_2$ are the $n \times r$ (vertical) loading matrices of the $n$ row objects and $n$ column objects on $(r < n)$ types of objects, respectively, and $\bF_t$ and $\bE_t$ are defined the same as in those in (\ref{eqn:fac_AA}). And this formulation is the matrix factor model considered in \cite{wang2016factor}. 

Model (\ref{eqn:fac_AA}) describes asymmetric relationships among objects in terms of asymmetric relationships among a single set of underlying components of the objects. Model (\ref{eqn:fac_A1A2}) is a more general mode where there are two sets of underlying components, and the directional relationships are hypothesized to hold from components of one kind to components of the other kind. In the international trade example, model (\ref{eqn:fac_AA}) would identify a single set of components or characteristics of countries given in the $\bA$ matrix of (\ref{eqn:fac_AA}), and provide $\bF_t$ matrices that describes the general effect of possessing any given characteristic on a country's tendency to trade with others with particular characteristics. An other way of putting this is to say that $F_t$ describe how much each type of country tend to trade with each of the other types. In contrast, model (\ref{eqn:fac_A1A2}) provides two sets of udnerlying components: $\bA_1$ gives the components of the objects in their row position and $\bA_2$ gives the components of the objects in their column position. The $\bF_t$ then gives the directed relationships from the $\bA_1$ components to the $\bA_2$ components. $\bA_1$ describes the amount of import-related factors possessed by countries and $\bA_2$ describes the amount of the export-related factors possessed by countries. And $\bF_t$ represents the amount that import-related factor generally prompts a country to import from a particular export-related factor. 

It can easily be shown that when $\bA_1$ and $\bA_2$ are not linear transformation of one another, then there generally exists no solution of the form given in (\ref{eqn:fac_AA}) for data fit by (\ref{eqn:fac_A1A2}) unless one goes to a higher dimensionality. Consequently, the model of equation (\ref{eqn:fac_AA}) makes a strong claim about a given data set. When the row dimensions and the column dimensions of a given asymmetrical relationship matrix can be demonstrated to span the same space, this agreement is a fact unlikely to arise by chanse and probably demonstrates the existe(\ref{eqn:fac_AA}). With data containing noise, of course, the row space and column space will probably not match exactly, but close agreement might still be interpreted as surprising the interesting. We will not discuss statistical tests of the fit of the two models in this article, but will demonstrate comparisons of the two models fit to a given set of real data in Section \ref{sec:application}.

The general method of interpreting model (\ref{eqn:fac_AA}) can be demonstrated by referring back to the example of international trade, where $X_{ij,t}$ expresses the trade flow from nation $i$ to nation $j$. For discussion, let's consider a $5$-dimension solution. The model (\ref{eqn:fac_AA}) would describe five basic factors or dimensions underlying the pattern of international trade behavior for these countries. These latent factors might be thought of as five idealized ``types'' of countries, types which each real country resembles to various degrees. This dimensions would be named by examining the loading matrix $\bA$ to see which individual countries have high loadings on each dimension. For example, if the individuals involved were members of major petroleum production countries, a label of ``petroleum abundant type'' might be assigned to the dimension on which petroleum production countries had high loadings. Other dimensions such as ``agriculture type'', ``high-tech type'' and ``active type'' might emerge from the analysis. Countries do not necessarily belong exclusively to a given ``type''. They can have moderate loadings on any given dimension and high loadings on more than one dimension. 
Alternatively, one might think of the dimensions as trading traits and the loadings as dexribing how much of each trait the country had in international trading. 

The $\bF_t$ matrix would not relate to specific countries, but instead provide a general statement of the patterns of trading among the five types of countries. Each row of the $\bF_t$ matrix would describe how much a given type of country generally tends to trade with the other four types or the same type if the diagonal element is considered. For example, $F_{ij,t}$ is the amount that ``agriculture type'' gnerally tend to import from ``high-tech type'', and $F_{ji,t}$ gives the converse relation. 

Alternatively, if the dimensions are thought of as country traits, each element of the $\bF_t$ matrix would describe how much the presence of a given country traits in a country generally tend to import from countries possessing another particular country trait.  

\section{Estimation Procedure and Properties}  \label{sec:est}

Similar to all factor models, the latent factors in the proposed model (\ref{eqn:fac_AA}) for asymmetric directional matrix time series can be linearly transformed into alternative factors with no loss of fit to the data. In general, if $\bH$ is any nonsingular $r \times r$ transformation matrix, we can define an alternative $\bA$ matrix, $\stackrel{\ast}{\bA}$, by letting $\stackrel{\ast}{\bA} = \bA \bH$ and defining the associated $\bF_t$ matrix $\stackrel{\ast}{\bF_t} = \bH^{-1} \bF_t \bH^{' -1}$. We may assume that the columns of $\bA$ are orthonormal, that is, $\bA'\bA=\bI_r$, where $\bI_r$ denotes the identity matrix of dimension $r$. Even with these constraints, $\bA$ and $\bF_t$ are not uniquely determined in (\ref{eqn:fac_AA}), as aforementioned linear transformation is still valid for any orthonormal $\bH$. However, the column space of the loading matrix $\bA$ is uniquely determined. Hence, in what follows, we will focus on the estimation of the column space of $\bA$. We denote the factor loading spaces by $\calM(\bA)$. For simplicity, we will depress the matrix column space notation and use the matrix notation directly. 


To facilitate the estimation, we use the QR decomposition $\bA=\bQ \bW$ to normalize the loading matrices, so that model (\ref{eqn:fac_AA}) can be re-expressed as 
\begin{equation}  \label{eqn:fac_AA_qr}
\bX_t = \bA \bF_t \bA' + \bE_t = \bQ \bZ_t \bQ' + \bE_t, \qquad t = 1, 2, \ldots, T,  
\end{equation}
where $\bZ_t = \bW \bF_t \bW'$ and $\bQ' \bQ = \bI_{r}$. 

Consider column vectors in (\ref{eqn:fac_AA_qr}), we write
\begin{equation}  \label{eqn:fac_AA_qr_col}
X_{t, \cdot j} = \bA \bF_t A_{j \cdot} + E_{t, \cdot j} = \bQ \bZ_t Q_{j \cdot} + E_{t, \cdot j}, \qquad j = 1, 2, \ldots, n, \;  t = 1, 2, \ldots, T.  
\end{equation}

We assume that both $\bF_t$ and $\bE_t$ are zero mean and thus $E(X_{t, \cdot j})=0$. Let $h$ be a positive integer. For $i,j = 1, 2, \ldots, n$, define 
\begin{align}  
& \bOmega_{zq, ij}(h) = \frac{1}{T-h} \sum_{t=1}^{T-h} Cov(\bZ_t Q_{i \cdot}, \bZ_{t+h} Q_{j \cdot})     \label{eqn:Omega_zqijh} \\
& \bOmega_{x, ij}(h) = \frac{1}{T-h} \sum_{t=1}^{T-h} Cov(X_{t, \cdot i}, X_{t+h, \cdot j}), \label{eqn:Omega_xijh}
\end{align}
which can be interpreted as the auto-cross-covariance matrices at lag $h$ between column $i$ and column $j$ of $\{\bZ_t \bQ'\}_{t=1, \cdots, T}$ and $\{\bX_t\}_{t=1, \cdots, T}$, respectively. 

For $h \ge 1$, it follows from (\ref{eqn:fac_AA_qr_col}), (\ref{eqn:Omega_zqijh}) and (\ref{eqn:Omega_xijh}) that 
\begin{equation}  \label{eqn:Omega_xijh_Omega_zqijh_relation}
\bOmega_{x, ij}(h) = \bQ \, \bOmega_{zq, ij}(h) \, \bQ'. 
\end{equation}

For a fixed $h_0 \ge 1$ satisfying Condition 2 in Section \ref{sec:theory} , define
\begin{equation}  \label{eqn:M_col_def}
\bM_{col} = \sum_{h=1}^{h_0} \sum_{i=1}^{n} \sum_{j=1}^{n} \bOmega_{x,ij}(h) \bOmega_{x,ij}(h)' = \bQ \left\{ \sum_{h=1}^{h_0} \sum_{i=1}^{n} \sum_{j=1}^{n} \bOmega_{zq, ij}(h) \bOmega_{zq, ij}(h)' \right\} \bQ'.  
\end{equation}

Similar to the column vector version, we define $\bM$ matrix for the row vectors of $\bX_t$'s as following
\begin{equation}  \label{eqn:M_row_def}
\bM_{row} = \sum_{h=1}^{h_0} \sum_{i=1}^{n} \sum_{j=1}^{n} \bOmega_{x',ij}(h) \bOmega_{x',ij}(h)' = \bQ \left\{ \sum_{h=1}^{h_0} \sum_{i=1}^{n} \sum_{j=1}^{n} \bOmega_{z'q, ij}(h) \bOmega_{z'q, ij}(h)' \right\} \bQ',  
\end{equation}
where $\bOmega_{z'q, ij}(h) = \frac{1}{T-h} \sum_{t=1}^{T-h} Cov(\bZ'_t Q_{i \cdot}, \bZ'_{t+h} Q_{j \cdot})$ and $\bOmega_{x', ij}(h) = \frac{1}{T-h} \sum_{t=1}^{T-h} Cov(X_{t, i \cdot}, X_{t, j \cdot})$. 

Finally, we define $\bM  =  \bM_{col} + \bM_{row}$, that is 
\begin{eqnarray}  
\bM & = & \sum_{h=1}^{h_0} \sum_{i=1}^{n} \sum_{j=1}^{n} \left( \bOmega_{x,ij}(h) \bOmega_{x,ij}(h)' + \bOmega_{x',ij}(h) \bOmega_{x',ij}(h)' \right) \nonumber \\
& = & \bQ \left\{ \sum_{h=1}^{h_0} \sum_{i=1}^{n} \sum_{j=1}^{n}  \left( \bOmega_{zq, ij}(h) \bOmega_{zq, ij}(h)' + \bOmega_{z'q, ij}(h) \bOmega_{z'q, ij}(h)' \right) \right\} \bQ'.  \label{eqn:M_def}
\end{eqnarray}

Obviously $\bM$ is a $n \times n$ non-negative definite matrix. Applying the spectral decomposition to the positive definite matrix sandwiched by $\bQ$ and $\bQ'$ on the right side of (\ref{eqn:M_def}), we have 
\begin{equation}  
\bM = \bQ \left\{ \sum_{h=1}^{h_0} \sum_{i=1}^{n} \sum_{j=1}^{n}  \left( \bOmega_{zq, ij}(h) \bOmega_{zq, ij}(h)' + \bOmega_{z'q, ij}(h) \bOmega_{z'q, ij}(h)' \right) \right\} \bQ' = \bQ \mathbf{U} \mathbf{D} \mathbf{U}' \bQ',  \nonumber
\end{equation}
where $\mathbf{U}$ is a $r \times r$ orthogonal matrix and $\mathbf{D}$ is a diagonal matrix with diagonal elements in descending order. As $\mathbf{U}'\bQ'\bQ\mathbf{U} = \bI_{r}$, the columns of $\bQ\mathbf{U}$ are the eigenvectors of $\bM$ corresponding to its $r$ non-zero eigenvalues. Thus the eigenspace of $\bM$ is the same as $\mathcal{M}(\bQ \mathbf{U})$ which is the same as $\mathcal{M}(\bQ)$. 
Under certain regularity conditions, the matrix $\bM$ has rank $r$. Hence, the columns of the factor loading matrix $\bQ$ can be estimated by the $r$ orthogonal eigenvectors of the matrix $\bM$ corresponding to its $r$ non-zero eigenvalues and the columns are arranged such that the corresponding eigenvalues are in the descending order. 

Now we define the sample versions of these quantities and introduce the estimation procedure. Suppose we have centered the observations $\{\bX_t\}_{t=1, \ldots, T}$, then for $h \ge 1$ and a prescribed positive integer $h_0$, let 
\begin{equation}  \label{eqn:Mhat_def}
\hat{\bM} = \sum_{h=1}^{h_0} \sum_{i=1}^{n} \sum_{j=1}^{n} \left( \hat{\bOmega}_{x,ij}(h) \hat{\bOmega}_{x,ij}(h)' + \hat{\bOmega}_{x',ij}(h) \hat{\bOmega}_{x',ij}(h)' \right), 
\end{equation}
where $\hat{\bOmega}_{x, ij}(h) = \frac{1}{T-h} \sum_{t=1}^{T-h} X_{t, \cdot i} X'_{t+h, \cdot j}$ and $\hat{\bOmega}_{x', ij}(h) = \frac{1}{T-h} \sum_{t=1}^{T-h} X_{t, i \cdot} X'_{t+h, j \cdot}$.

A natural estimator for the $\bQ$ specified above is defined as $\hat{\bQ} = \left\{ \hat{\mathbf{q}}, \cdots, \hat{\mathbf{q}}_{r} \right\}$, where $\hat{\mathbf{q}}_i$ is the eigenvector of $\hat{\bM}$ corresponding to its $i$-th largest eigenvalue. Consequently, we estimate the factors and residuals respectively by 
\begin{equation}
\hat{\bZ}_t = \hat{\bQ}' \bX_t \hat{\bQ}, \quad \text{and} \quad \hat{\bE}_t = \bX_t - \hat{\bQ} \hat{\bZ}_t \hat{\bQ}' = (\bI_{n} - \hat{\bQ} \hat{\bQ}') \bX_t + \hat{\bQ} \hat{\bQ}' \bX_t (\bI_{n} - \hat{\bQ} \hat{\bQ}'). 
\end{equation}

The above estimation procedure assumes the number of row factors $r$ is known. To determine $r$ we could use the eigenvalue ratio-based estimator of \cite{lam2012factor}.  Let $\hat{\lambda}_1 \ge \hat{\lambda}_2 \ge \cdots \ge \hat{\lambda}_{r} \ge 0$ be the ordered eigenvalues of $\hat{\bM}$. The ratio-based estimator for $r$ is defined as 
\begin{equation}  \label{eqn:eigen_ratio_r}
\hat{r} = \arg \min_{1 \le j \le r_{\max}} \frac{\hat{\lambda}_{j+1}}{\hat{\lambda}_j}, 
\end{equation}
where $r \le r_{\max} \le n$ is an integer. In practice we may take $r_{\max} = n/2$ or $r_{\max} = n/3$. 


\section{Simulations} \label{sec:simu}

In this section, we use simulation to study the performance of the estimationn methods in Section \ref{sec:est}. In the simulations, the observed data ${\bX_t}_{t=1:T}$ are generated according to model (\ref{eqn:fac_AA}),
\begin{equation} 
	\bX_t = \bA \bF_t \bA' + \bE_t, t = 1, 2, \ldots, T. \nonumber
\end{equation}
We choose the dimensions of the latent network $\bF_t$ to be $r=3$. The entries of $\bF_t$ follow $r^2$ independent AR(1) processes with Gaussian while noise $\calN(0,1)$ innocations. Specifically, $vec(\bF_t)=\bPhi_F vec(\bF_{t-1}) + \bepsilon_t$ with $\bPhi_F=diag(0.86, 0.93, 0.81, 0.73, 0.62, 0.61, 0.53, 0.75, 0.78)$. The entries of $\bA$ are independently sampled from uniform distribution $U(-p^{-\delta/2}, p^{-\delta/2})$ and the factor strength is controlled by parameter $\delta$. The disturbance $\bE_t$ is a white noise process with mean zero and a Kronecker product covariance structure, that is, $Cov(vec(\bE_t)) = \bGamma_2 \otimes \bGamma_1$, where $\bGamma_1$ and $\bGamma_2$ are both of sized $p \times p$. Both $\bGamma_1$ and $\bGamma_2$ have values 1 on the diagonal and 0.2 on the off-diagonal entries. 

We first study the performance of our proposed approach on estimating the loading spaces. Table \ref{table:A_MSD} shows the results for estimating the loading spaces $\calM(\bA)$. The accuracies are measured by the space distance using the correct dimension of the latent network, that is $r=3$. Estimators $\hat{\bA}_R$, $\hat{\bA}_C$ and $\hat{\bA}_{RnC}$ are estimated from $\bM_{row}$, $\bM_{col}$ and $\bM$, respectively. The results show that with stronger signals and more data sample points, the estimation accuracy increases. Moreover, estimator from the combination of row and column information $\bM$ is the best among three in the sense that it is the closest to the truth. 

\begin{table}[htpb!]
\centering
\resizebox{\textwidth}{!}{
\begin{tabular}{cc|ccc|ccc|ccc|ccc}
\hline
 &  & \multicolumn{3}{c|}{$T = 0.5 \, n^2$} & \multicolumn{3}{c|}{$T = n^2$} & \multicolumn{3}{c|}{$T = 1.5 \, n^2$} & \multicolumn{3}{c}{$T = 2 \, n^2$} \\ \hline
$\delta$ & $n$ & $\mathcal{D}(\hat{A}_R, A)$ & $\mathcal{D}(\hat{A}_C, A)$ & $\mathcal{D}(\hat{A}_{RnC}, A)$ & $\mathcal{D}(\hat{A}_R, A)$ & $\mathcal{D}(\hat{A}_C, A)$ & $\mathcal{D}(\hat{A}_{RnC}, A)$ & $\mathcal{D}(\hat{A}_R, A)$ & $\mathcal{D}(\hat{A}_C, A)$ & $\mathcal{D}(\hat{A}_{RnC}, A)$ & $\mathcal{D}(\hat{A}_R, A)$ & $\mathcal{D}(\hat{A}_C, A)$ & $\mathcal{D}(\hat{A}_{RnC}, A)$ \\ \hline
0 & 20 & 0.27(0.05) & 0.46(0.14) & 0.21(0.04) & 0.17(0.03) & 0.21(0.04) & 0.12(0.02) & 0.08(0.01) & 0.12(0.02) & 0.06(0.01) & 0.11(0.02) & 0.14(0.02) & 0.08(0.01) \\
0 & 40 & 0.08(0.01) & 0.10(0.01) & 0.06(0.01) & 0.06(0.01) & 0.07(0.01) & 0.04(0.01) & 0.04(0.00) & 0.05(0.01) & 0.03(0.00) & 0.04(0.00) & 0.05(0.01) & 0.03(0.00) \\
0 & 60 & 0.03(0.00) & 0.05(0.01) & 0.03(0.00) & 0.03(0.00) & 0.04(0.00) & 0.02(0.00) & 0.02(0.00) & 0.03(0.00) & 0.02(0.00) & 0.02(0.00) & 0.03(0.00) & 0.02(0.00) \\ \hline
0.5 & 20 & 5.54(0.08) & 5.76(0.05) & 5.61(0.07) & 5.08(0.27) & 5.60(0.07) & 4.81(1.03) & 1.30(0.51) & 4.98(0.21) & 1.24(0.55) & 2.39(0.38) & 3.49(0.22) & 2.43(0.29) \\
0.5 & 40 & 5.70(0.07) & 5.65(0.07) & 5.59(0.04) & 5.54(0.11) & 5.13(0.08) & 5.50(0.15) & 4.18(0.93) & 5.66(0.09) & 4.22(0.6) & 5.70(0.16) & 5.76(0.03) & 5.75(0.02) \\
0.5 & 60 & 5.69(0.06) & 5.49(0.03) & 5.71(0.02) & 2.79(0.61) & 5.6(0.02) & 2.50(0.91) & 5.11(0.17) & 5.59(0.02) & 5.19(0.1) & 4.71(0.66) & 5.08(0.05) & 4.88(0.22) \\ \hline
\end{tabular}
} 
\caption{Means and standard deviations (in parentheses) of the estimation accuracy measured by $\mathcal{D}(\hat{A}, A)$. For ease of presentation, all numbers in this table are the true numbers multiplied by $10$. The results are average of 200 simulations. }
\label{table:A_MSD}
\end{table}

Now we present the performance of our proposed approach on estimating the dimension of the latent network $r=3$. In table \ref{table:correct_r_freq}, $f_R$, $f_C$, and $f_RnC$ represents the frequency of correctly estimating the dimension using $\bM_{row}$, $\bM_{col}$ and $\bM$, respectively. Again, the results show that with stronger signals and more data sample points, the estimation accuracy increases. Moreover, estimator from the combination of row and column information $\bM$ is the best among three in the sense that it has the highest frequency of correctly estimating the number of latent dimensions.     

\begin{table}[htpb!]
\centering
\resizebox{\textwidth}{!}{
\begin{tabular}{cc|ccc|ccc|ccc|ccc}
\hline
 &  & \multicolumn{3}{c|}{$T = 0.5 \, n^2$} & \multicolumn{3}{c|}{$T = n^2$} & \multicolumn{3}{c|}{$T = 1.5 \, n^2$} & \multicolumn{3}{c}{$T = 2 \, n^2$} \\ \hline
$\delta$ & $n$ & $f_R$ & $f_C$ & $f_{RnC}$ & $f_R$ & $f_C$ & $f_{RnC}$ & $f_R$ & $f_C$ & $f_{RnC}$ & $f_R$ & $f_C$ & $f_{RnC}$ \\ \hline
0 & 20 & 0.975 & 0.3 & 0.965 & 0.99 & 0.72 & 1 & 1 & 1 & 1 & 1 & 1 & 1 \\
0 & 40 & 1 & 1 & 1 & 1 & 1 & 1 & 1 & 1 & 1 & 1 & 1 & 1 \\
0 & 60 & 1 & 1 & 1 & 1 & 1 & 1 & 1 & 1 & 1 & 1 & 1 & 1 \\ \hline
0.5 & 20 & 0.72 & 0.085 & 0.805 & 0.345 & 0.3 & 0 & 0 & 0 & 0 & 0.005 & 0.005 & 0.21 \\
0.5 & 40 & 0 & 0 & 0.06 & 0 & 0 & 0 & 0 & 0 & 0 & 0 & 0 & 0 \\
0.5 & 60 & 0 & 0 & 0 & 0 & 0 & 0 & 0 & 0 & 0 & 0 & 0 & 0 \\ \hline
\end{tabular}
}  
\caption{Relative frequencies of correctly estimating the dimension of the latent network. The results are based on 200 simulations.}
\label{table:correct_r_freq}
\end{table}

\section{Application to International Trade Flow Data}  \label{sec:application}

\subsection{International Trade Flow Data}
International Monetary Fund (IMF) publishes \emph{Direction of Trade Statistics} (DOTS) (\cite{IMF-DOTS}) that provides monthly data on the country and area distribution of countries' exports and imports by their partners.
We focus on aggregated monthly trading volumes of a $420$-month period from January, 1981 to December, 2015 of all goods denominated in U.S. dollars.
In the following analysis, only import data are fed into our factor model because it is generally believed that they are more accurate than export ones (\cite{durand1953countryclassification}; \cite{linnemann1966econometric}).
This is especially true when we are interested in tracing countries of production and consumption rather than countries of consignment or of purchase and sale (\cite{linnemann1966econometric}). 



Even though DOTS provides data for all $420$ months for $105$ countries, the quality of data vary across time and countries.
Some countries failed to report their volumes of trade in some or all years.
After removing countries with missing values, we select following $23$ countries for constructing our time series: Australia, Canada, China Mainland, Denmark, Finland, France, Germany, Hong Kong, Indonesia, Ireland, Italy, Japan, Korea, Malaysia, Mexico, Netherlands, New Zealand, Singapore, Spain, Sweden, Thailand, United Kingdom, United States.

The length of our dynamic network matrices time series is $420$, as we do not remove any monthly data from analysis.
At each time, the observation is a square matrix of $24$ rows and columns, where each row or column corresponds to a country.
Each cell in the matrix contains the dollar trade volume that the country corresponding to its row imports from the country corresponding to its column.
Hence rows represent import relationship and columns represent export relationship. 

By examining the network of trade flows, we will show in the following text that we can analyze how countries compare to each other in terms of trade volumes and patterns and how these volumes and patterns evolve as economical cycles and political events unfold.
We want to emphasize that our analysis does not draw on aggregate country statistics such as GNP, production statistics or any other national attributes. 


\subsection{Five-Year Rolling Estimation} \label{subsec:5_year_rolling}

To ease our presentation, we break the $420$-month period into $31$ rolling $5$-year periods: $1981$ through $1985$, $1982$ through $1986$ and so forth.
For each $5$-year period, we assume that the loadings are constant and estimate the loading matrix $\bA$ under model (\ref{eqn:fac_AA}) and $\bA_1$ and $\bA_2$ under model (\ref{eqn:fac_A1A2}).
Fix the number of factors $r$, for each of the $31$ periods, we estimate 3 loading matrices $\bA$, $\bA_1$ and $\bA_2$, whose dimensions are $24 \times r$.
We index these matrices by the mid-year of the five-year periods.
For example, $\bA$ for period $1981$--$1985$ is indexed with year $1983$, $\bA$ for period $1982$--$1986$ is indexed with year $1984$ and so forth.    

The diagonals for these loading matrices are undefined, and will be ignored in the analysis by estimating their value from the model in an interative procedure, identical to that sometimes used for communality estimation in factor analysis. 

Which $\bH$ we select can depend on which perspective we wish to take toward the interpretation of $\bA$ and $\bF_t$. Although in general we might like to seek some kind of approximate simple structure for the columns of $\bA$, this can be done in different ways, corresponding roughly to different orthogonal or oblique rotation criteria in factor analysis. 

In this article, we will adopt as standard a procedure which applies Varimax to the columns of $\bA$ \textit{after} they have been scaled to have equal sums of sqaures; this keeps the columns of $\bA$ mutually orthogonal.

When the columns of $A$ are standardized, the $\bF_t$ matrix can be interpreted as expressing relationships among the latent factors in the same units as the original data. That is, the $\bF_t$ matrices can be interpreted as a matrix of the same kind as the original data matrices $\bX_t$, but describing the relations among the latent aspects of the objects, rather than the objects themselves. 

In the analyses presented in this article, we have adopted the convention that the columns of $\bA$ are standardized so that they sum to $1.0$. This is feasible because we are dealing with data which contain all positive values, and our columns of $\bA$ will contain mostly positive entries. With other data, we might adopt the convention that the sum of squares for each column of $\bA$ sums to $1.0$, which is obviously feasible regardless of the sign pattern of the data or of the entries in $\bA$.

When the columns of $\bA$ are standardized to have sums equal to $1.0$, then the $\bF_t$ matrices can be thought of as a compressed or miniature version of the original $\bX_t$ matrix. The sum of all the elements in $\bF_t$ is equal to the sum of all elements in $\hat{\bX}_t$, the part of $\bX_t$ fit by the model. 

\subsection{Results on Trading Level}  \label{subsec:res_on_level} 

\begin{figure}[ht!]
	\centering
	\includegraphics[width=\textwidth, keepaspectratio]{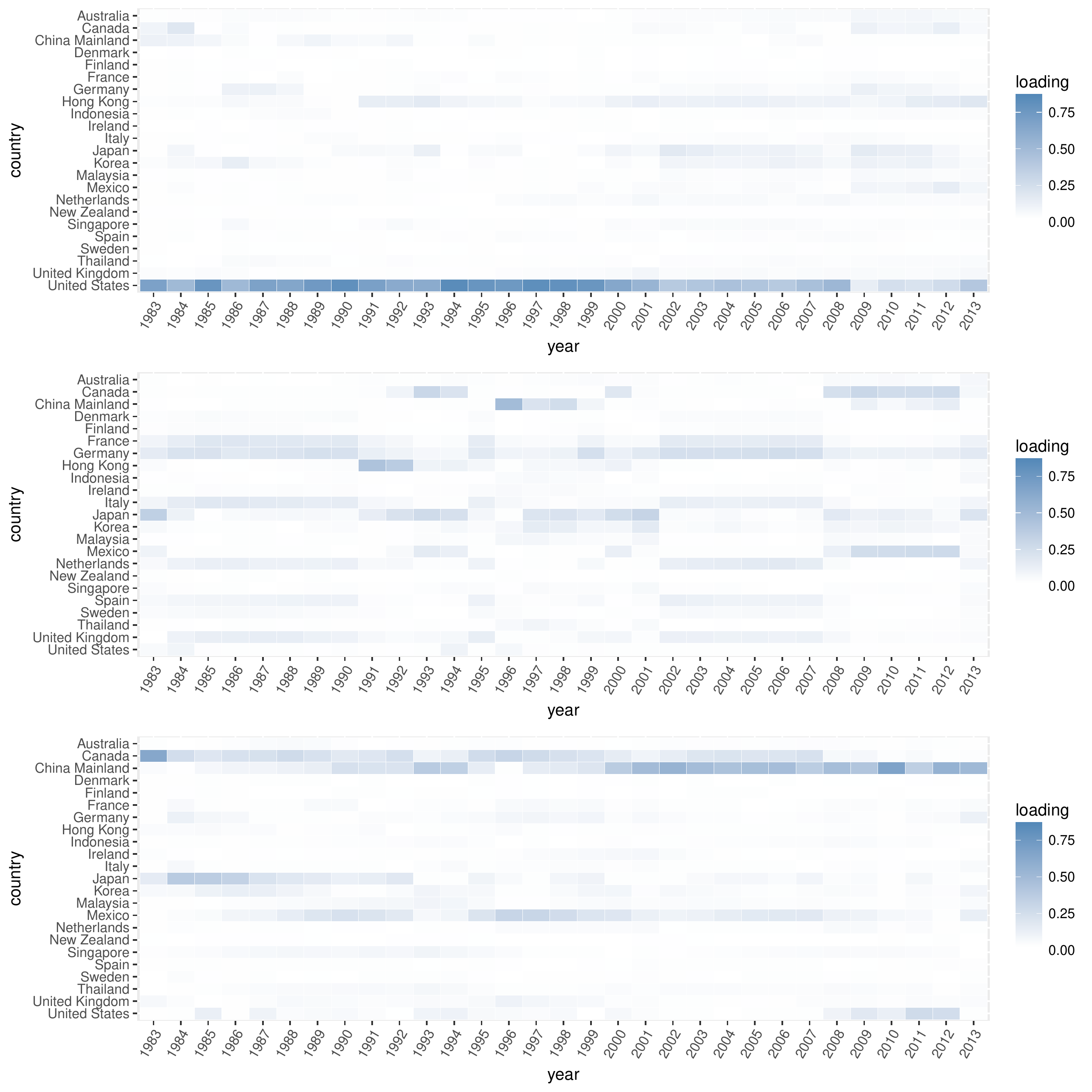}
	\caption{Latent factor loadings for trading level on $r = 3$ dimensions for a series of $31$ rolling five-year periods indexed from $1983$ to $2013$. Top plot contains factors loaded maximum on the United States. Middle plot contains factors loaded maximum on Germany. Bottom plot contains the remaining factors. The factors loaded maximum on the United States and Germany are distinct for all periods.}
	\label{fig:AA_dyn_3_fac_all}
\end{figure}

We apply the methods described in Section \ref{sec:est} and Section \ref{subsec:5_year_rolling} to the level of international trade flow data. For illustration, we use $r=3$ factors. The estimated loading matrix $\hat{\bA}$ has been rotated by Varimax approximation to simple structures, and its columns are kept orthonormal. We then scale each column of $\hat{\bA}$ such that its sum is $1$. Figure \ref{fig:AA_dyn_3_fac_all} presents the heatmaps of the loadings from $1983$ to $2013$. Three vertically aligned heatmaps correspond to three columns of loading matrix $\hat{\bA}$ from year $1983$ to $2013$, respectively. The columns of $\hat{\bA}$ from different years are aligned according to their maximum loading on the United States and Germany for the top and middle plots. There is no overlapping of factors on their maximum loading on United States and Germany. The bottom plot contains the remaining factor for all the years. There are negative values in estimated loadings $\hat{\bA}$. However, they are very close to zero and occur rarely, thus we set all negative values to zeros.   

The factors in the first plot can be interpreted as representing the United States, as the loading of the United States on this dimension dominates all other countries. The factors in the second plot are aligned according to the maximum loading on Germany, and not surprisingly, they are also heavily loaded on European countries such as France, Italy, Netherlands, Spain and United Kingdom. Therefore, this dimension can be interpreted as representing European countries. We should also note that Japan's loading on this dimension is also significant, which suggests that Japan's international trading activity is similar to that of those European countries. The third factor features sizable loadings on Canada, China Mainland, Japan and Mexico. Thus the third dimension represents a group of large economies besides US and Europe. 

The evolution in the loadings of the countries on the first three dimensions over these $31$ years is striking. Recall that each column in a heat map sums up to $1.0$. Thus, the value at each cell denotes a country's participation in a dimension at a certain year. Figure \ref{fig:AA_dyn_3_fac_all} (a) shows that the United States dominates the first dimension from $1983$ to $2013$. However, its participation in the first dimension gradually decreases since $2002$ and reaches its minimal in year $2009$ and $2010$, signaling the aftermath of the $2008$ financial crisis. The decrease from United States is offset by increase from Canada, China, Hong Kong, Japan, Korea and Mexico, which is manifested by the increasingly darker cells since $2002$ for those countries.    

Figure \ref{fig:AA_dyn_3_fac_all} (b) presents the evolution of the second European dimension. Before $1986$, trading activities in the European group are dominated by France, Germany, and Italy. From $1987$ onwards, France, Germany, Italy, Netherlands, Spain, and United Kingdom all participate actively in trade while France, Germany and Italy still take a major part. With the introduction of Euro in $2002$, Netherlands, Spain, and United Kingdom's participations in trade increase. However, other countries such as China, Canada, Japan and Mexico also participate heavily at times on this dimension.

Figure \ref{fig:AA_dyn_3_fac_all} (c) shows the evolution of the third dimension of international trade besides the US and Europe. Before $1988$, Canada and Japan are the dominating countries in this group. Mexico starts taking a large portion around $1994$ when the North American Free Trade Agreement (NAFTA) took effect. China's activities grow gradually from moderate in $1987$ to dominating in the period from $2000$ to $2013$.  



\begin{figure}[ht!]
	\centering
	\includegraphics[width=\textwidth, keepaspectratio]{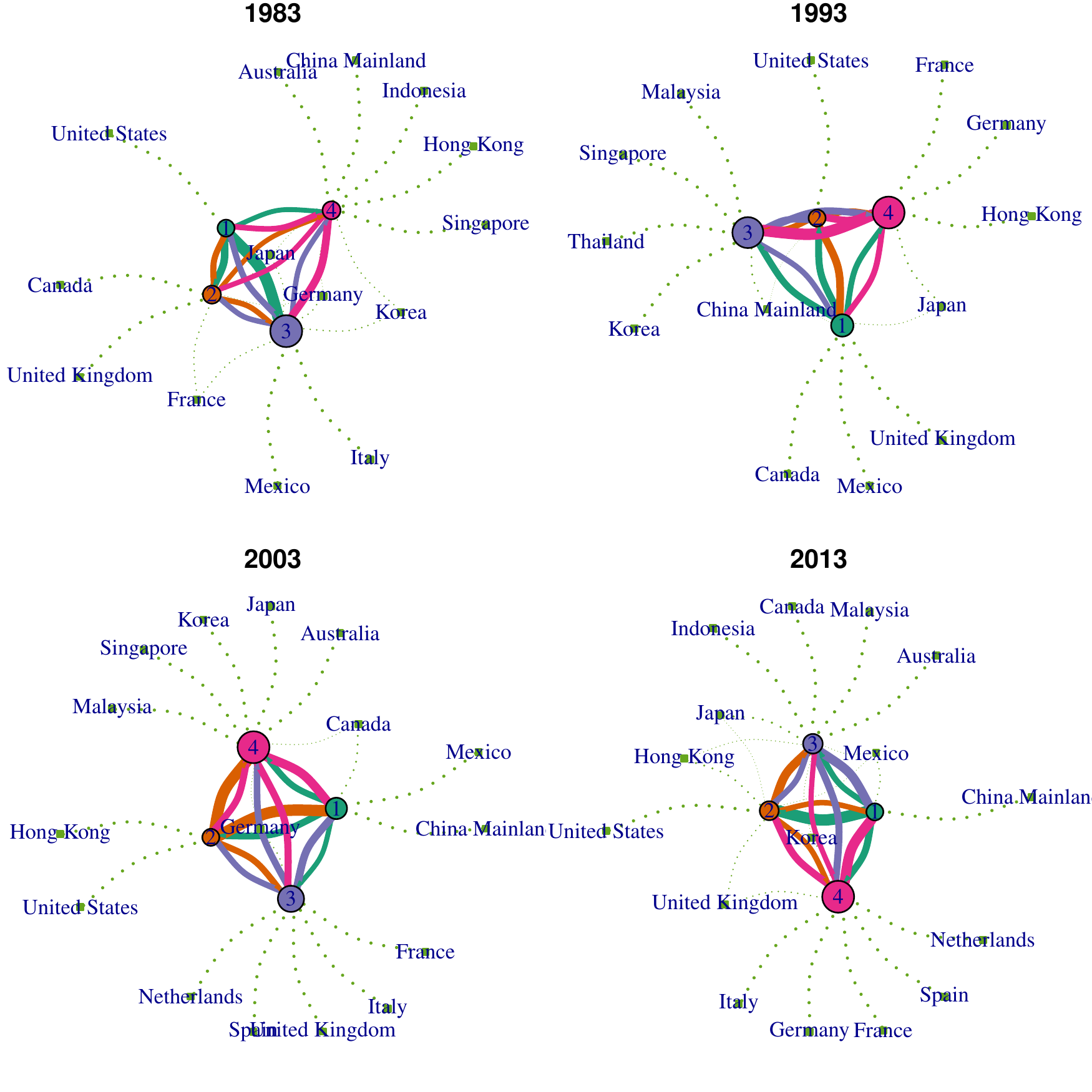}
	\caption{Trading level network plot of latent dimensions and relationship between countries and the latent dimensions. Thickness of the solid line represents the volume of trades among latent dimensions. Thickness of the dotted lines represents the level of connection between latent dimensions and countries. Note that a country can be related to multiple latent dimensions.}
	\label{fig:raw_network_combined_rr_4_3}
\end{figure}

Figure \ref{fig:raw_network_combined_rr_4_3} plots the trading network among $4$ latent dimensions as well as the relationship between countries and latent dimensions for $4$ selected years. The trading network among latent dimensions is plotted based on the $4 \times 4$ latent factor matrix. The colored circles represent $4$ latent dimensions. The size of each circle conveys the trading volumes within each latent dimension, i.e., the values of the diagonal elements in the latent factor matrix. The width of the solid lines connecting circles conveys the trading volume between different latent dimensions, i.e., the values of the off-diagonal elements in the latent factor matrix. The direction of the flow is conveyed by the color of the line. Specifically, the color of the line is the same as its import dimension. For example, a red line connecting a blue dimension and a red one represents the trade flow from the blue factor to the red one. Note that the widths of the solid lines across different network plots are not comparable because they are scaled to fit each individual plot, otherwise the lines in the 2013 plot will overwhelm the whole plot because the trading volume is much larger in 2013.


The relationships between countries and $4$ latent dimensions, shown as the dotted lines, are plotted using a simplified version of the estimated loading matrix $\hat{\bA}$ to provide an uncluttered view. Specifically, we generate a base matrix by rounding $10 \hat{\bA}$. We set all non-dominating entries to zero for each row (country) of the base matrix, and then re-weight the non-zero entries such the sum of row is $1$. We alternate between these $2$ steps until no changes occur. The non-dominating entries for each row are defined as values that are more than $0.5$ smaller than the maximum entry of the row. The countries with zero loadings in the resulting matrix are not plotted. The size of the dotted line conveys the strength of connection between a country and a latent dimension.   

Clearly shown in the network evolution, in the 80's the United States stood out as a single dimension. The second dimension is composed of Canada, France, Germany and United Kingdom. The third dimension is composed of France, Germany, Italy, Japan, Korea and Mexico. The fourth dimension is composed of Australia, China, Germany, Hong Kong, Indonesia, Japan, Korea and Singapore.


As shown by the thick green lines, the first dimension, representing the U.S., imports mostly from the second and the third dimensions, which load mostly on European and America countries. The countries other than the North America and European ones form the fourth dimension, which import a small amount from the other three dimensions as shown by the thin pink line. The trading within this dimension is small too. Note that Japan and Germany load on three dimensions. 


In 1993, the United States still formed a single dimension -- the second dimension in orange. The third dimension in purple is dominated then by China, Korea, Malaysia, Singapore and Thailand, which were the quickly developing countries in Asia during the 80's and early 90's. Thus, dimension $3$ can be interpreted as the Asian dimension. Dimension $1$ is composed of Canada, Mexico, United Kingdom. Dimension $4$ is composed of France, Germany, Hong Kong and Japan. Note that Japan loads on both dimension $1$ and $4$, but it loads more on $4$. As shown by the thick pink line connecting dimension $3$ and $4$, dimension $4$ (Europe group?) imported a lot from the Asian dimension. Also, its import from the United States dimension is sizable. The United States dimension imports more from dimension $1$ (the America dimension) than from other dimensions. 

In 2003, dimension $1$, composed of Canada, China and Mexico, represents the dimension that exports a lot to dimension $2$ (Hong Kong and United States). Dimension $4$ is composed of Australia, Japan, Korea, Malaysia and Singapore. France, Italy, Netherlands, Spain and United Kingdom make up dimension $3$, which can be interpreted as the European dimension. The European dimension becomes tighter than those in 1993 and 1983 because the Euro was introduced in 2002. 

In 2013, China dominates a single dimension. The European dimension, which are composed of France, Italy, Germany, Netherlands, Spain and United Kingdom, does not change from 2003. However, the within Europe trading volumes (the size of pink dimension $4$) increase a lot when compared with that in 2003. The United States still loaded completely on dimension $2$. However, Hong Kong and United Kingdom also load partially on dimension $2$, indicating that these two countries share some similar import/export pattern as the US. For example, Hong Kong and United Kingdom export a lot to the China dimension (the thick green line between dimension $1$ and $2$) and import a lot from the Asian dimension (the thick orange line between dimension $2$ and $3$).  

\begin{figure}[ht!]
	\centering
	\includegraphics[width=\textwidth, keepaspectratio]{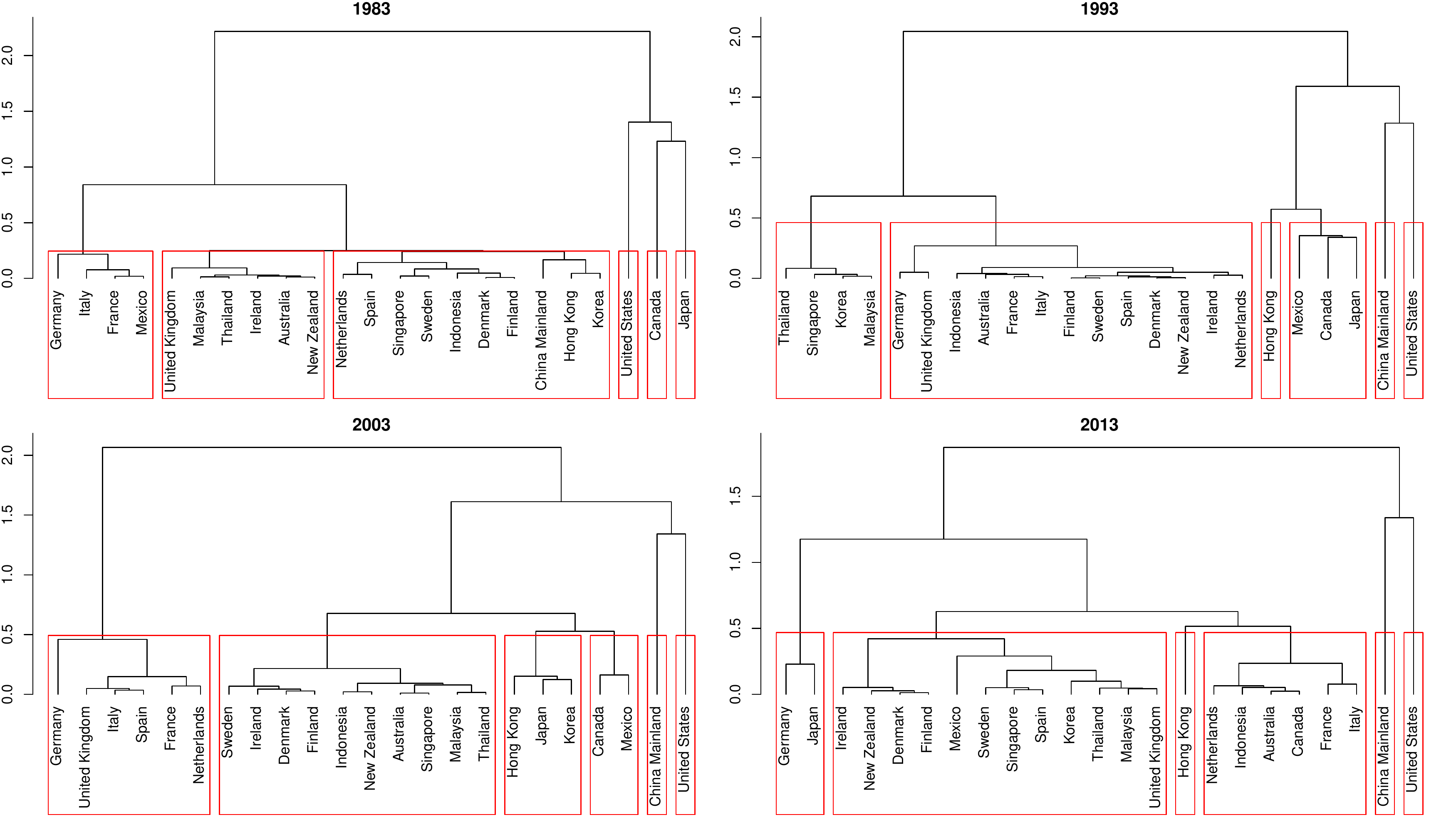}
	\caption{Clustering of countries based on their trading level latent dimension representations.}
	\label{fig:raw_hclust_fac_3_four_plots_yr_3}
\end{figure}

Figure \ref{fig:raw_hclust_fac_3_four_plots_yr_3} shows the clustering of countries based on their loadings on first four latent dimensions over years. The rectangles denotes clusters that divide countries into six groups. It offers a new perspective to inspect the dynamics of countries' trading behaviors. The United States accounts for a single cluster for all years because of its large trading volumes with other countries. China's weight in the global trade over the years has been gradually increasing: in 1983 China's trading behavior is more like Asian economies such as Hong Kong and Korea. However, from 1990's to 2010's, as China's trade becomes more active, its trading behavior becomes more similar to that of the United States and it makes up single cluster.  

\subsection{Results on Trading Growth}  \label{subsec:res_on_growth} 

To investigate the evolution of international trade growth pattern from 1981 to 2015, we apply the proposed method to log differenced data. Since we don't expect a country's import growth pattern to be the same as its export growth pattern, we use model (\ref{eqn:fac_A1A2}) with different right and left loading matrices $\bA_1$ and $\bA_2$. The estimation and plotting methods are the same as those described in Section \ref{subsec:5_year_rolling} and Section \ref{subsec:res_on_level}. Note that for the log differenced data, the leading dimensions in latent matrix factor explain the most variance in trade growth, not absolute trade levels. 

\begin{figure}[ht!]
	\centering
	\includegraphics[width=\textwidth, keepaspectratio]{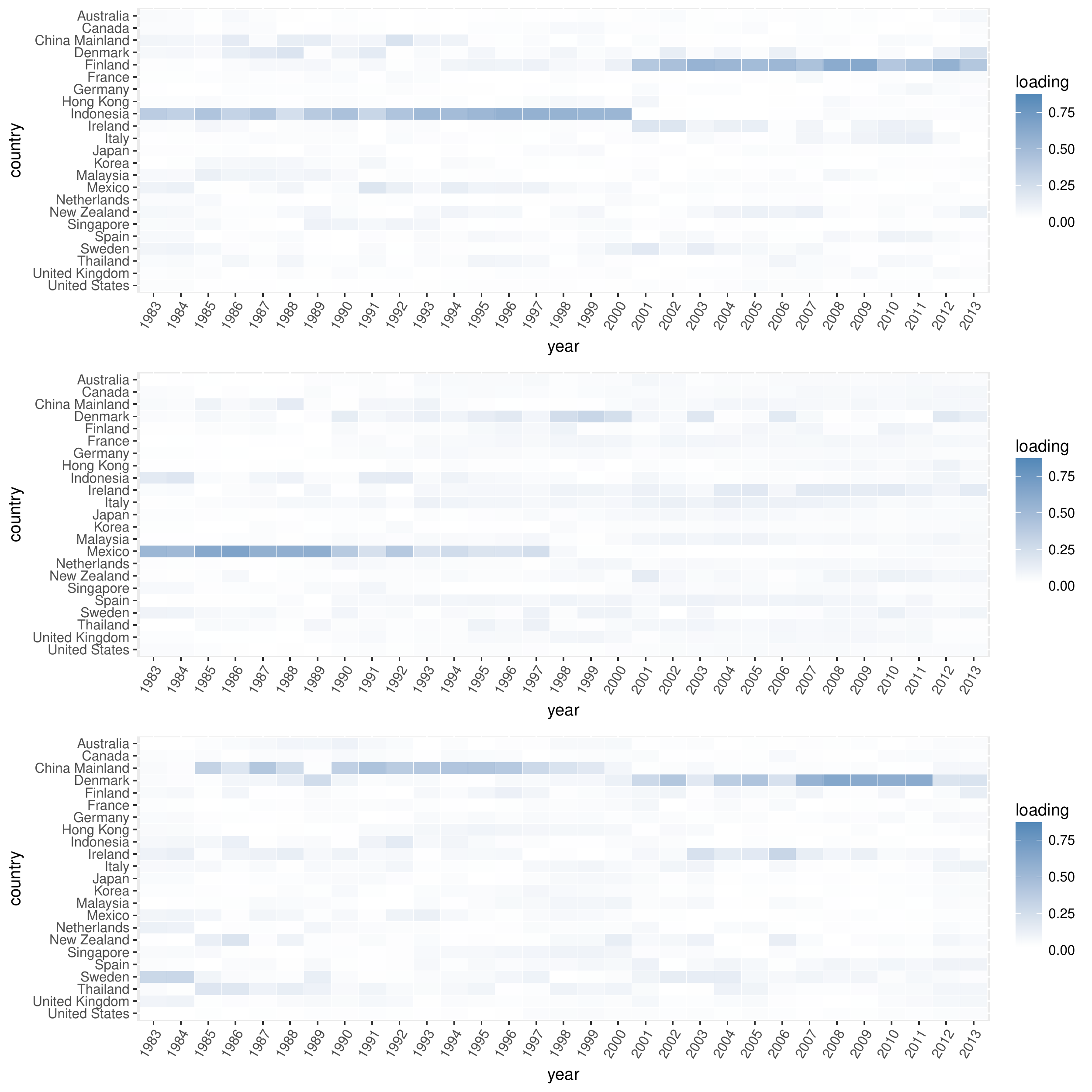}
	\caption{Import factor loadings for trading growth on $r = 3$ dimensions for a series of $31$ rolling five-year periods indexed from $1983$ to $2013$. Top plot contains factors loaded maximum on the Indonesia. Middle plot contains factors loaded maximum on Mexico. Bottom plot contains the remaining factors. The factors loaded maximum on the Indonesia and Mexico are distinct for all periods.}
	\label{fig:dln_import_dyn_3_fac_all}
\end{figure}

Figure \ref{fig:dln_import_dyn_3_fac_all} plots rotated import factor loadings on $r = 3$ dimensions for a series of $31$ rolling five-year periods indexed from $1983$ to $2013$. Each plot corresponds to a row (import) dimension of the latent matrix factor. Indonesia loads heavily on the first dimension till 2000. After 2000, Finland loads most on the first dimension. Mexico loads heavily on the second dimension till 1990. After 1990, Denmark loads slightly more than Mexico on this dimension. After 2004, Ireland starts to show the largest loading on the second dimension. But all other countries also shares closely on the same dimension. 
China loads heavily on the third dimension from 1985 to 2000. Denmark and Ireland load heavily on the third dimension after 2001 when Euro was introduced. 

\begin{figure}[ht!]
	\centering
	\includegraphics[width=\textwidth, keepaspectratio]{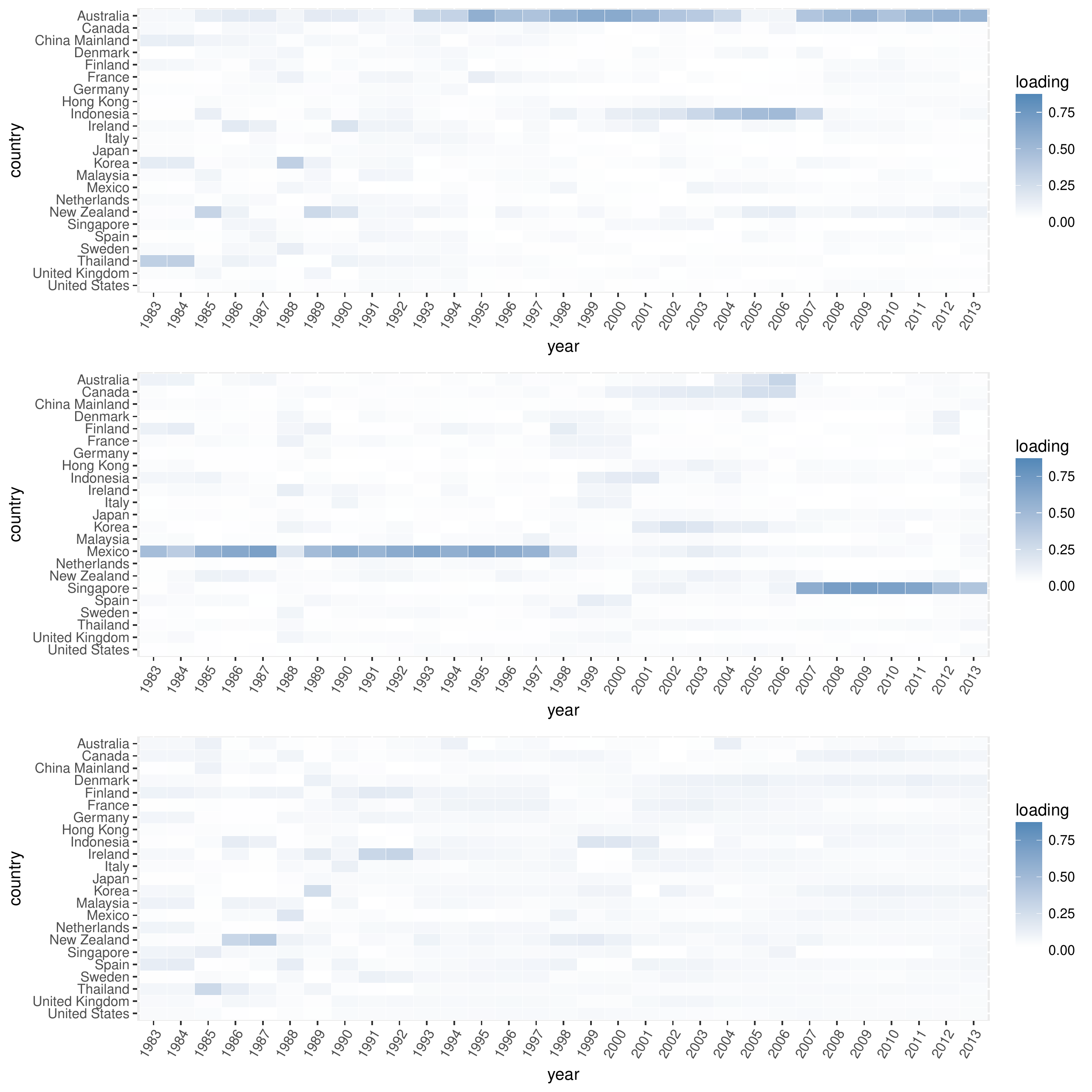}
	\caption{Export factor loadings for trading growth on $r = 3$ dimensions for a series of $31$ rolling five-year periods indexed from $1983$ to $2013$. Top plot contains factors loaded maximum on the Australia. Middle plot contains factors loaded maximum on Mexico. Bottom plot contains the remaining factors. The factors loaded maximum on the Australia and Mexico are distinct for all periods.}
	\label{fig:dln_export_dyn_3_fac_all}
\end{figure}

Figure \ref{fig:dln_export_dyn_3_fac_all} plots export factor loadings on $r = 3$ dimensions for a series of $31$ rolling five-year periods indexed from $1983$ to $2013$. Each plot corresponds to a column (export) dimension of the latent matrix factor. 

The export growth is very different from the import growth. Australia's export growth loads the most on the first dimension of latent matrix factor all the years except from 2001 to 2007 when Indonesia becomes prominent on this dimension. 
Mexico's export growth loads the most on the second dimension of latent matrix factor model from 1983 to 1998 and Singapore loads the most from 2006 to 2013. From 1998 to 2006, Australia, Canada, Finland, France, Germany, Hong Kong, Indonesia, Korea, Malaysia, Mexico, New Zealand, Singapore and Spain load lightly on the second dimension. 
The third dimension does not feature prominent countries like the first and second. 

\begin{figure}[ht!]
	\centering
	\includegraphics[width=\textwidth, keepaspectratio]{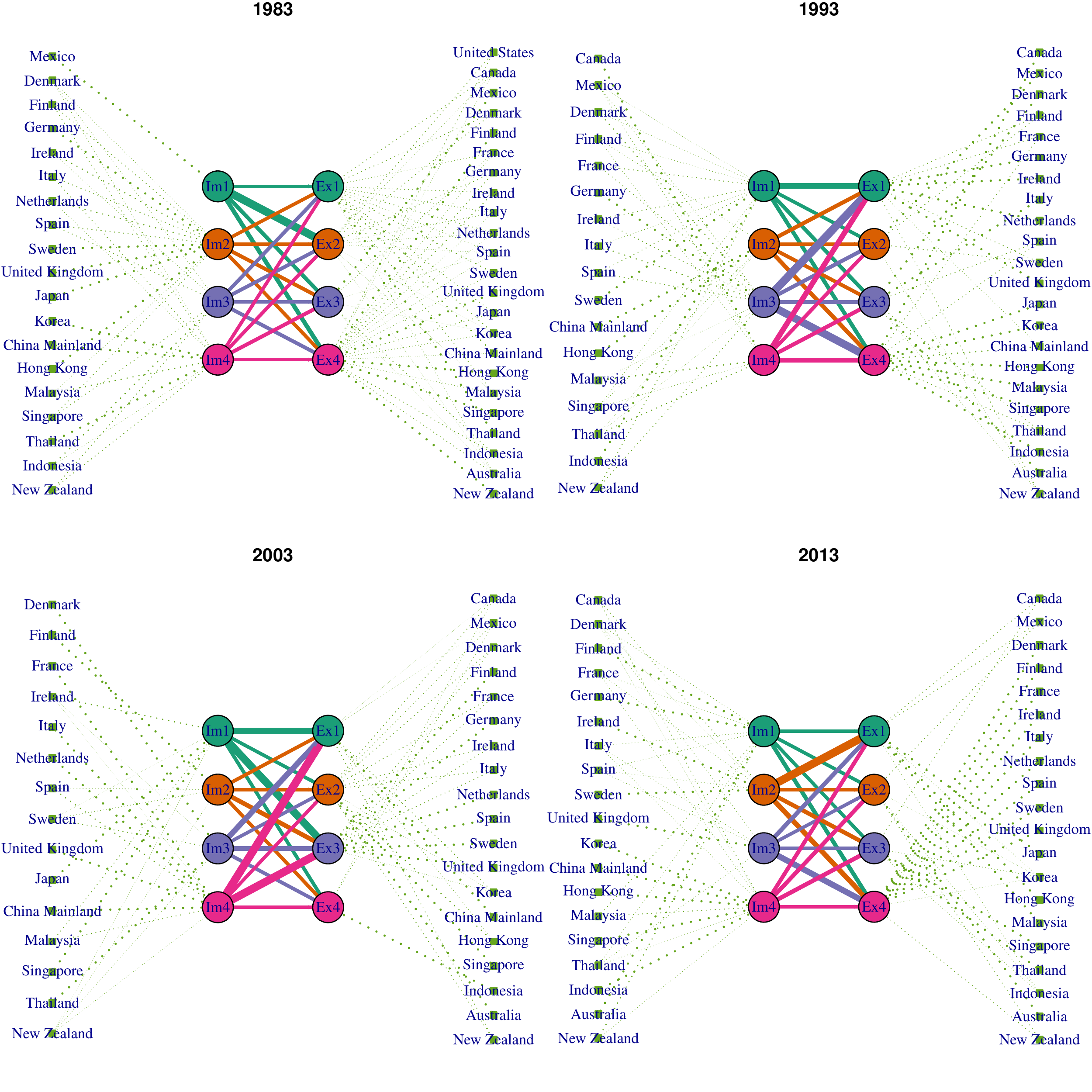}
	\caption{Trading growth network plot of latent dimensions and relationship between countries and the latent dimensions. Thickness of the solid line represents the volume of trades among latent dimensions. Thickness of the dotted lines represents the level of connection between latent dimensions and countries. Note that a country can be related to multiple latent dimensions.}
	\label{fig:dln_network_combined_rr_4_3}
\end{figure}

Figure \ref{fig:dln_network_combined_rr_4_3} plots the trading network among import and export latent dimensions as well as the relationships between countries and latent dimensions for $4$ selcted years. Since we use different left and right loading matrix, the relationships between countries and latent dimensions are different for import and export activities. And the meaning of the row and column dimensions of the latent matrix factor are different. Specifically, the row dimensions represent the import groups or characteristics while the column dimensions correspond to the export counterparts. Thus in Figure \ref{fig:dln_network_combined_rr_4_3}, we distinguish the row and column dimensions and have eight circle nodes for the latent dimensions. The nodes with prefix ``Im'' and ``Ex'' in their notations correspond to the import (row) dimensions and the export (column) dimensions, respectively.  

In 1983 most trading growth happened between import latent dimension $1$ and the export latent dimension $2$, as shown by the thick green line connecting them. The import latent dimension $1$ is mostly composed of Mexico and thus represents Mexico import activities. The export latent dimension $2$ is composed mostly of Canada, Denmark, Germany, Italy, Netherland, Spain, United States and slightly of Australia, Korea and thus can be interpreted as a mixture of developed economies. The trading growths between other import and export dimensions seem to be very similar magnitudes. The meaning of the import and export latent dimensions are apparent from the plot too. The import latent dimension $2$ and $4$ clearly represent the European and the Asia countries, respectively. The import latent dimension $3$ composes of Denmark, Indonesia, Ireland, New Zealand, Spain and Singapore. Mexico loads solely on the export latent dimension $3$, which can be interpreted as the export group/characteristics of Mexico. Export latent dimension $1$ and $4$ have most connections with a mixture of advanced economies from North America, Europe, and Asian. 

In 1993 most trading growth happened between import latent dimension $3$ and the export latent dimension $1$ as well as between import latent dimension $3$ and the export latent dimension $4$, as shown by the two thick purple lines connecting both pairs. Clearly, the import latent dimension $3$ stands for the European countries due to its strong connection with all European countries. The export latent dimension $1$ is a mixture of European countries, United States, China and New Zealand, although it concentrates more on Europe and United States. It is also very clear that the export latent dimension $4$ represents Asian countries. The connections show that in 1993 most trading growth happened within European and between European and Asian countries.

In 2003 most trading growth happened between import latent dimension $4$ and the export latent dimension $1$ and $3$, as shown by the two thick pink lines connecting both pairs. And the trading growth between import latent dimension $1$ and export latent dimension $1$ and $3$ are also prominent, as shown by the two thick green lines connecting both pairs. As shown by the thick dotted line, Japan and Singapore loaded heavily on the import latent dimension $1$. Thus the import latent dimension $1$ stands for developed Asian countries. The export latent dimension $1$ connects mostly to Asia and Oceania countries, thus representing these groups. And the export dimension $3$ represents the European countries as evidenced by its connections. Thus, around 2003, imports of developed Asian economies, such as Japan and Singapore, from Asian, Oceania, and European countries grows noticeably. The import latent dimension $4$ represents both European countries and China Mainland. The thick pink lines from the import latent dimension $4$ to export latent dimension $1$ and $3$ reveal that import growth of European and China from European, Asian, and Oceania countries are notable. The grouping effect among European countries is more obvious in 2003 than previous years since clearly the important latent dimension $4$ and export latent dimension $3$ represent European countries. This manifests the increase of European trade after January 1994 when the European Economic Area (EEA) was formed to provide for the free movement of persons, goods, services and capital within the internal market of the European Union as well as $3$ of the $4$ member states of the European Free Trade Association. And in January 2002 $12$ countries of the European Union launched the Euro zone (euro in cash), which instantly became the second most used currency in the world. 

In 2013, most trading growth happened between import latent dimension $2$ ``Im2'' and the export latent dimension $1$ ``Ex1'', as shown by the thick orange lines connecting the pair. The import dimension $1$ represents a mixture of European and Asian countries. The import dimension $2$ represents European countries with emphasis on Ireland and Sweden. The import dimension $3$ also represents European countries but with an emphasis on Unite Kingdom. The import dimension $4$ represents Asian countries as it is mostly connected to them. The meaning of export latent dimension $1$ is much clearer. The export latent dimension $1$ strongly connects to Japan, Hong Kong and Korea, thus represents developed Asia economies. The export latent dimenion $2$ represents Oceania countries because it strongly connects to Australia and New Zealand. The export latent dimenion $3$ can be interpreted as Southeast Asia countries because Indonesia, Singapore and Thailand have high loadings. The export latent dimension $4$ ``Ex4'' can be labeled European, since all European countries have high loadings. 


\begin{figure}[ht!]
	\centering
	\includegraphics[height=\textheight, keepaspectratio]{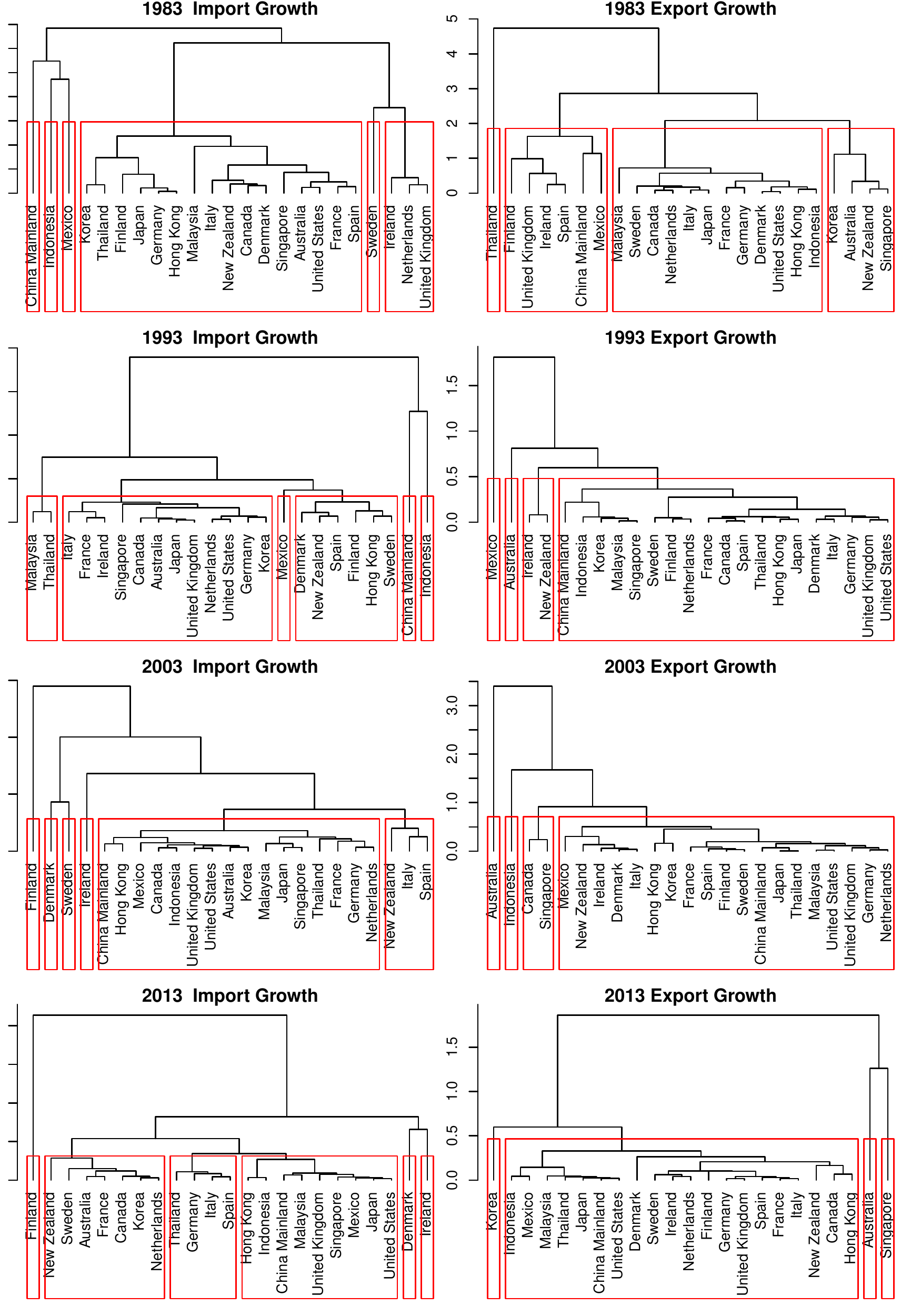}
	\caption{Clustering of countries based on their trading growth latent dimension representations.}
	\label{fig:dln_dendrogram_combined_rr_4_3}
\end{figure}


\newpage
\bibliographystyle{chicago}
\bibliography{\mybib}

\end{document}